# Operator-Schmidt decomposition and the geometrical edges of two-qubit gates

S. Balakrishnan · R. Sankaranarayanan

**Abstract**   Nonlocal two-qubit quantum gates are represented by canonical decomposition or equivalently by operator-Schmidt decomposition. The former decomposition results in geometrical representation such that all the two-qubit gates form tetrahedron within which perfect entanglers form a polyhedron. On the other hand, it is known from the later decomposition that Schmidt number of nonlocal gates can be either 2 or 4. In this work, some aspects of later decomposition are investigated. It is shown that two gates differing by local operations possess same set of Schmidt coefficients. Employing geometrical method, it is established that Schmidt number 2 corresponds to controlled unitary gates. Further, all the edges of tetrahedron and polyhedron are characterized using Schmidt strength, a measure of operator entanglement. It is found that one edge of the tetrahedron possesses the maximum Schmidt strength, implying that all the gates in the edge are maximally entangled.

S. Balakrishnan[1*,]   R. Sankaranarayanan[2]

Department of Physics, National Institute of Technology, Tiruchirappalli 620015, India

1. physicsbalki@gmail.com
2. sankar@nitt.edu
    *corresponding author

# 1 Introduction

Entanglement is a nonlocal property of quantum states, playing central role in quantum information processing [1, 2]. For a two-qubit system, entanglement can be controlled by appropriate application of nonlocal quantum unitary operators (gates). It is then important to characterize and classify the nonlocal attributes of two-qubit gates. One important characterizing tool is local invariants of a gate, which are unaffected by local operations [3]. Zhang *et al*. studied quantum gates from geometrical perspective using canonical decomposition [4]. The local invariants and geometrical approach compliments each other, such that nonlocal two-qubit gates form an irreducible geometry of tetrahedron (known as Weyl chamber). Of all the gates, exactly half of them are perfect entanglers – capable of producing maximally entangled state when acting on some separable states. Geometrically, the perfect entanglers form a polyhedron within the Weyl chamber [4].

Another useful representation of two-qubit operator is operator-Schmidt decomposition. In this representation, the number of non zero (Schmidt) coefficients $s_l$ of a gate is called as Schmidt number. It is known that local gates have Schmidt number 1 and nonlocal gates have Schmidt number 2 or 4 [5, 6]. In this paper, geometrical method is employed to show that Schmidt number 2 belong to the controlled unitary gates; implying that controlled-NOT is the *only* perfect entangler possessing Schmidt number 2. Further, it is shown that gates possessing same local invariants must necessarily have the *same* set of $s_l$.

To capture the entangling capabilities of unitary operators, Zanardi introduced the notion of operator entanglement through linear entropy $L(U)$. It is known that $L(U)$ is related to the entangling power, which is defined as the average entanglement production of an operator [7, 8]. Another less studied measure of operator entanglement is the Schmidt strength $K_{Sch}(U)$, which is defined as the Shannon entropy of $s_l^2$ associated to a gate [5]. D. Collins *et al*. introduced a framework to measure the nonlocal content of a gate, in terms of resources (entangled and classical bits) required for implementation of the gate using double teleportation. Within this framework, SWAP and Double-CNOT are known to be maximally nonlocal gates [9, 10]. Here, we found that one edge of the tetrahedron, which includes SWAP and Double-CNOT, possesses the maximum Schmidt strength. It implies that all the gates in the edge are maximally entangled. This result naturally enquires if all the other gates in the edge are maximally nonlocal as well, in the sense of resources required for their implementation using double teleportation.

The paper is organized as follows. In the next section, canonical decomposition of two-qubit gates and associated geometry is briefed. From the operator-Schmidt decomposition, it is shown in Section 3 that gates differing by local operations must have same set of coefficients $s_l$. A simple geometrical analysis is presented to show that Schmidt number 2 corresponds to controlled unitary gates. In Section 4, coefficients $s_l$ for all the edges of tetrahedron and polyhedron are computed, from which the Schmidt strength is analyzed. The paper is concluded with a discussion on the results presented.

## 2  Canonical decomposition

An arbitrary two-qubit gate $U \in \mathrm{SU}(4)$ can be written in the following form known as canonical decomposition [11, 12]:

$$U = k_1 \exp\left\{\frac{i}{2}(c_1 \sigma_x^1 \sigma_x^2 + c_2 \sigma_y^1 \sigma_y^2 + c_3 \sigma_z^1 \sigma_z^2)\right\} k_2 \tag{1}$$

where $\sigma_x, \sigma_y, \sigma_z$ are Pauli matrices and $k_1, k_2 \in \mathrm{SU}(2) \otimes \mathrm{SU}(2)$. Two unitary operators $U, U_1 \in \mathrm{SU}(4)$ are called locally equivalent if they differ only by local operations: $U = k_1 U_1 k_2$. A class of gates differ from $U$ only by local operations is referred as local equivalence class $[U]$. Makhlin and Zhang *et al*. have shown that the local equivalence class $[U]$ can be characterized *uniquely* by local invariants which are calculated as follows. Representing $U$ in the Bell basis,

$$\left|\Phi^+\right\rangle = \frac{1}{\sqrt{2}}(\left|00\right\rangle + \left|11\right\rangle), \quad \left|\Phi^-\right\rangle = \frac{i}{\sqrt{2}}(\left|01\right\rangle + \left|10\right\rangle),$$

$$\left|\Psi^+\right\rangle = \frac{1}{\sqrt{2}}(\left|01\right\rangle - \left|10\right\rangle), \quad \left|\Psi^-\right\rangle = \frac{i}{\sqrt{2}}(\left|00\right\rangle - \left|11\right\rangle)$$

as $U_B = Q^\dagger U Q$, with

$$Q = \frac{1}{\sqrt{2}}\begin{pmatrix} 1 & 0 & 0 & i \\ 0 & i & 1 & 0 \\ 0 & i & -1 & 0 \\ 1 & 0 & 0 & -i \end{pmatrix},$$

local invariants are defined as [3,4]

$$G_1 = \frac{tr^2[M(U)]}{16\det(U)}, \tag{2a}$$

$$G_2 = \frac{tr^2[M(U)] - tr[M^2(U)]}{4\det(U)} \tag{2b}$$

where $M(U) = U_B^T U_B$. Local invariants and a point $[c_1, c_2, c_3]$ corresponding to the gate $U$ are related as [4]

$$G_1 = \cos^2 c_1 \cos^2 c_2 \cos^2 c_3 - \sin^2 c_1 \sin^2 c_2 \sin^2 c_3 + \frac{i}{4}\sin 2c_1 \sin 2c_2 \sin 2c_3, \tag{3a}$$

$$G_2 = 4\cos^2 c_1 \cos^2 c_2 \cos^2 c_3 - 4\sin^2 c_1 \sin^2 c_2 \sin^2 c_3 - \cos 2c_1 \cos 2c_2 \cos 2c_3. \tag{3b}$$

From this relation, for given local invariants $(G_1, G_2)$ the point $[c_1, c_2, c_3]$ in a 3-torus geometry (with period $\pi$) is identified. In other words, the gate $U$ or its equivalence class $[U]$ is characterized by the point $[c_1, c_2, c_3]$ as well. The symmetry reduced 3-torus takes the form of tetrahedron (Weyl chamber). A two-qubit gate is called a perfect entangler if it produces a maximally entangled state when acting on some separable input state. Perfect entanglers constitute a polyhedron within the Weyl chamber as shown in Fig.1.

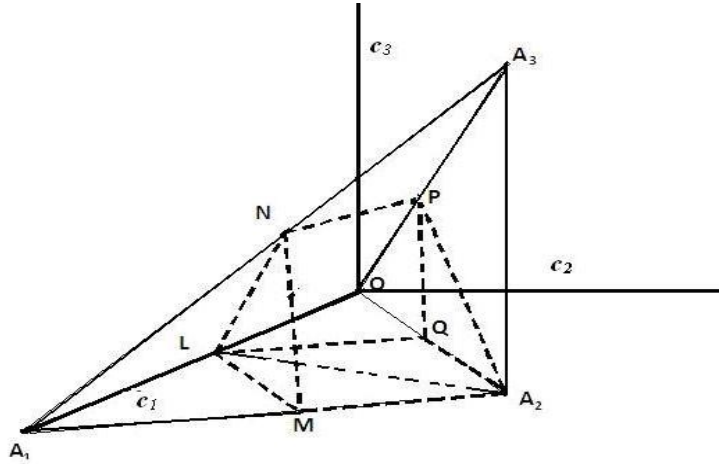

**Fig. 1** Tetrahedron $OA_1A_2A_3$ (Weyl chamber) is the geometrical representation of nonlocal two-qubit gates. Polyhedron $LMNPQA_2$ (shown in dashed lines) corresponds to the perfect entanglers. The thick lines represent the $c_1$, $c_2$ and $c_3$ axes of the Weyl chamber. The points $L$, $M$, $N$, $P$, and $Q$ are midpoints of the tetrahedron edges $OA_1$, $A_2A_1$, $A_1A_3$, $OA_3$, and $OA_2$ respectively. The points $L = [\pi/2, 0, 0]$, $A_2 = [\pi/2, \pi/2, 0]$ and $A_3 = [\pi/2, \pi/2, \pi/2]$ correspond to CNOT, Double-CNOT and SWAP gates respectively

## 3 Operator-Schmidt decomposition

An alternate representation to canonical decomposition is the operator-Schmidt decomposition using which an operator $U$ can be expressed as

$$U = \sum_l s_l\, A_l \otimes B_l \qquad (4)$$

where $s_l \geq 0$ are called as Schmidt coefficients and $A_l$ ($B_l$) are orthonormal operator bases for system $A$ ($B$) [5]. In this representation, the number of non-zero Schmidt coefficients of an operator is defined as *Schmidt number*. Leaving the local operations $k_1$ and $k_2$, nonlocal content of $U$ as given in Eq. 1 can be rewritten in the operator-Schmidt decomposition as follows. Since $\sigma_x \otimes \sigma_x$, $\sigma_y \otimes \sigma_y$ and $\sigma_z \otimes \sigma_z$ commute each other, Eq. 1 can be written as

$$U = \exp\left(\frac{i}{2} c_1 \sigma_x \otimes \sigma_x\right) \exp\left(\frac{i}{2} c_2 \sigma_y \otimes \sigma_y\right) \exp\left(\frac{i}{2} c_3 \sigma_z \otimes \sigma_z\right).$$

Since $(\sigma_x \otimes \sigma_x)^2 = (\sigma_y \otimes \sigma_y)^2 = (\sigma_z \otimes \sigma_z)^2 = I$, we have the following relations

$$\exp\left(\frac{i}{2} c_1 \sigma_x \otimes \sigma_x\right) = I \otimes I \cos\left(\frac{c_1}{2}\right) + i \sigma_x \otimes \sigma_x \sin\left(\frac{c_1}{2}\right)$$

$$\exp\left(\frac{i}{2} c_2 \sigma_y \otimes \sigma_y\right) = I \otimes I \cos\left(\frac{c_2}{2}\right) + i \sigma_y \otimes \sigma_y \sin\left(\frac{c_2}{2}\right)$$

$$\exp\left(\frac{i}{2} c_3 \sigma_z \otimes \sigma_z\right) = I \otimes I \cos\left(\frac{c_3}{2}\right) + i \sigma_z \otimes \sigma_z \sin\left(\frac{c_3}{2}\right).$$

With this the nonlocal content of $U$ can be expressed as [5]

$$U = (C_1 C_2 C_3 + i S_1 S_2 S_3) I \otimes I + (C_1 S_2 S_3 + i S_1 C_2 C_3) \sigma_x \otimes \sigma_x$$
$$+ (S_1 C_2 S_3 + i C_1 S_2 C_3) \sigma_y \otimes \sigma_y + (S_1 S_2 C_3 + i C_1 C_2 S_3) \sigma_z \otimes \sigma_z$$

where $C_j = \cos(c_j/2)$ and $S_j = \sin(c_j/2)$.

Rewriting the above expression as

$$U = z_1 (I \otimes I) + z_2 (\sigma_x \otimes \sigma_x) + z_3 (\sigma_y \otimes \sigma_y) + z_4 (\sigma_z \otimes \sigma_z) \qquad (5)$$

where

$$z_1 = \frac{1}{2}\left[e^{ic_1/2}\cos\left(\frac{c_3-c_2}{2}\right) + e^{-ic_1/2}\cos\left(\frac{c_3+c_2}{2}\right)\right]$$

$$z_2 = \frac{1}{2}\left[e^{ic_1/2}\cos\left(\frac{c_3-c_2}{2}\right) - e^{-ic_1/2}\cos\left(\frac{c_3+c_2}{2}\right)\right] \quad (6)$$

$$z_3 = \frac{-i}{2}\left[e^{ic_1/2}\sin\left(\frac{c_3-c_2}{2}\right) - e^{-ic_1/2}\sin\left(\frac{c_3+c_2}{2}\right)\right]$$

$$z_4 = \frac{i}{2}\left[e^{ic_1/2}\sin\left(\frac{c_3-c_2}{2}\right) + e^{-ic_1/2}\sin\left(\frac{c_3+c_2}{2}\right)\right].$$

We observe that Eq. 5 is in the Schmidt form such that $|z_l|$ is recognized as Schmidt coefficients $s_l$. Representing Eq. 5 in the Bell basis and using Eq. 2, after some algebra, local invariants are related to the above coefficients as

$$G_1 = \left[\sum_{l=1}^{4} z_l^2\right]^2, \quad (7a)$$

$$G_2 = G_1 + 2\sum_{l=1}^{4} z_l^4 + 24\prod_{l=1}^{4} z_l. \quad (7b)$$

This relation enables to compute local invariants of a gate from the operator-Schmidt decomposition as well. From the above relation, it is clear that any permutation of $z_l$ will leave the local invariants unaffected. We may note that Eq. 7 will retain its form under the transformations: $z_l \to -z_l$, $z_l \to iz_l$, $z_l \to -iz_l$. Further, the local invariants are unaltered if $z_l \to \pm iz_l$ with two of the coefficients having same sign of transformation (to preserve the last term in Eq. 7b). Thus, local equivalence class of a gate can also be recognized from the coefficients $z_l$. In other words, locally equivalent gates possess *same* set of Schmidt coefficients $s_l$. However, the gates with same set of $s_l$ need not be locally equivalent.

While local gates possess Schmidt number 1, nonlocal gates correspond to Schmidt number 2 or 4. It may be noted that the gates can not possess Schmidt number 3 [5]. In what follows, we employ geometrical method to classify the nonlocal gates using Schmidt number. For Schmidt number 2 gates, if any one of the coefficients in Eq. 6 is zero for certain values of $c_1, c_2$ and $c_3$ then another coefficient also vanishes. The coefficients $z_l$ can be zero if (A) sum of the terms is zero or (B) individual terms are zero. Then the geometrical points for each vanishing coefficient are

$z_1 = 0$:  (A) $[0, c_2, \pi]$ or $[0, \pi, c_3]$  (B) $[c_1, 0, \pi]$

$z_2 = 0$:  (A) $[0, c_2, 0]$ or $[0, 0, c_3]$  (B) $[c_1, 0, \pi]$

$z_3 = 0$:  (A) $[0, c_2, \pi]$ or $[0, 0, c_3]$  (B) $[c_1, 0, 0]$

$z_4 = 0$:  (A) $[0, c_2, 0]$ or $[0, \pi, c_3]$  (B) $[c_1, 0, 0]$.

From these conditions, it is also clear that quantum gates can not have Schmidt number 3. For example, if $z_1 = 0$ then $z_3$ or $z_4$ or $z_2$ also vanishes. Hence, the possible gates with Schmidt number 2 are $[c_1, 0, \pi]$, $[0, c_2, \pi]$, $[0, \pi, c_3]$, $[c_1, 0, 0]$, $[0, c_2, 0]$ and $[0, 0, c_3]$. Since $c_i$ are arbitrary here, they are conveniently denoted as $\theta$ with $0 \leq \theta \leq \pi$. From the geometry, it is known that if $[c_1, c_2, c_3]$ is an element in a local equivalence class $[U]$ then $[c_i, c_j, c_k]$, $[\pi - c_i, \pi - c_j, c_k]$, $[\pi - c_i, c_j, \pi - c_k]$, and $[c_i, \pi - c_j, \pi - c_k]$ are also in $[U]$ where $(i, j, k)$ is a permutation of $(1, 2, 3)$. Hence, the first three possible values of $[c_1, c_2, c_3]$ corresponding to Schmidt number 2 are the permutations of $[\theta, \pi, 0]$, and the remaining three values are the permutations of $[\theta, 0, 0]$. Using $[\pi - c_i, \pi - c_j, c_k]$ we identify $[\theta, \pi, 0]$ as $[\pi - \theta, 0, 0]$. Therefore, we identify that the two lines $[\pi - \theta, 0, 0]$ and $[\theta, 0, 0]$ correspond to the gates with Schmidt number 2. The line $[\theta, 0, 0]$ corresponds to the edge $OA_1$ of the Weyl chamber (Fig. 1.), which is the well known controlled unitary gates; $[\pi - \theta, 0, 0]$ is the mirror image of $[\theta, 0, 0]$ and hence they are locally equivalent to each other. Therefore, it is clear that *only* controlled unitary gates $[\theta, 0, 0]$ with $0 \leq \theta \leq \pi/2$ possess Schmidt number 2 and all other nonlocal gates possess Schmidt number 4. Hence, Schmidt number 2 class is a special case of nonlocal gates which are outnumbered by Schmidt number 4 gates. Since CNOT is the only perfect entangler in the controlled unitary family, it is the only perfect entangler with Schmidt number 2.

It is known that a gate with Schmidt number 2 can be written in the following form upto the local equivalence

$$U = \sqrt{(1-p)}\, I \otimes I + i\sqrt{p}\, \sigma_x \otimes \sigma_x \tag{8}$$

where $0 \leq p \leq 1$ [5]. Substituting $p = \sin^2(\theta/2)$ we have

$$U = \cos\left(\frac{\theta}{2}\right) I \otimes I + i \sin\left(\frac{\theta}{2}\right) \sigma_x \otimes \sigma_x. \tag{9}$$

Using Eq. 7 the local invariants of the above operator is computed as $G_1 = \cos^2\theta$ and $G_2 = 2G_1 + 1$, which precisely correspond to the controlled unitary gates [4]. It is worth mentioning that $p$ and the geometrical parameter $\theta$ have one-to-one correspondence in the defined range.

## 4 Schmidt strength

Since the Schmidt coefficients satisfy $\sum_{l=1}^{4} s_l^2 = 1$, $s_l^2$ form a probability distribution. Exploiting this property, Schmidt strength is defined as the Shannon entropy of the distribution $s_l^2$:

$$K_{Sch}(U) = -\sum_{l=1}^{4} s_l^2 \log_2 s_l^2 \qquad (10)$$

which may recognized as a measure of entanglement of $U$ [5]. It is easy to check that $0 \leq K_{Sch} \leq 2$, such that $K_{Sch} = 0$ for local gates. In this section we compute Schmidt coefficients and hence Schmidt strength for six edges of tetrahedron (Weyl chamber) and nine edges of polyhedron (see Fig. 1). We may note that the geometry of all the edges are one parametric [13]. Subsequently, the Schmidt coefficients and Schmidt strength of the edges are also one parametric. By substituting geometrical points of the edges in Eq. 6, Schmidt coefficients $s_l$ are computed and presented in tables 1 and 2.

Table. 1 Schmidt coefficients for the edges of tetrahedron (Weyl chamber)

| Edge | $[c_1, c_2, c_3]$ | Range of parameter | $s_1$ | $s_2$ | $s_3$ | $s_4$ |
|---|---|---|---|---|---|---|
| $OA_1$ | $[\theta, 0, 0]$ | $0 \leq \theta \leq \pi$ | $\cos\left(\dfrac{\theta}{2}\right)$ | $\sin\left(\dfrac{\theta}{2}\right)$ | - | - |
| $OA_2$ | $[\theta, \theta, 0]$ | $0 \leq \theta \leq \dfrac{\pi}{2}$ | $\cos^2\left(\dfrac{\theta}{2}\right)$ | $\dfrac{1}{2}\sin\theta$ | $\dfrac{1}{2}\sin\theta$ | $\sin^2\left(\dfrac{\theta}{2}\right)$ |
| $A_2A_1$ | $\left[\dfrac{\pi}{2}+\phi, \dfrac{\pi}{2}-\phi, 0\right]$ | $0 \leq \phi \leq \dfrac{\pi}{2}$ | $\dfrac{1}{2}\cos\phi$ | $\dfrac{1}{2}[1+\sin\phi]$ | $\dfrac{1}{2}[1-\sin\phi]$ | $\dfrac{1}{2}\cos\phi$ |
| $A_2A_3$ | $\left[\dfrac{\pi}{2}, \dfrac{\pi}{2}, \phi\right]$ | $0 \leq \phi \leq \dfrac{\pi}{2}$ | $\dfrac{1}{2}$ | $\dfrac{1}{2}$ | $\dfrac{1}{2}$ | $\dfrac{1}{2}$ |
| $OA_3$ | $\left[\dfrac{\pi\alpha}{2}, \dfrac{\pi\alpha}{2}, \dfrac{\pi\alpha}{2}\right]$ | $0 \leq \alpha \leq 1$ | $\dfrac{1}{2}\sqrt{1+3\cos^2\left(\dfrac{\pi\alpha}{2}\right)}$ | $\dfrac{1}{2}\sin\left(\dfrac{\pi\alpha}{2}\right)$ | $\dfrac{1}{2}\sin\left(\dfrac{\pi\alpha}{2}\right)$ | $\dfrac{1}{2}\sin\left(\dfrac{\pi\alpha}{2}\right)$ |
| $A_1A_3$ | $\left[\pi-\dfrac{\pi\alpha}{2}, \dfrac{\pi\alpha}{2}, \dfrac{\pi\alpha}{2}\right]$ | $0 \leq \alpha \leq 1$ | $\dfrac{1}{2}\sin\left(\dfrac{\pi\alpha}{2}\right)$ | $\dfrac{1}{2}\sqrt{1+3\cos^2\left(\dfrac{\pi\alpha}{2}\right)}$ | $\dfrac{1}{2}\sin\left(\dfrac{\pi\alpha}{2}\right)$ | $\dfrac{1}{2}\sin\left(\dfrac{\pi\alpha}{2}\right)$ |

It is known from our earlier studies on the Weyl chamber that the edges $OA_3$ and $A_1A_3$ correspond to SWAP$^\alpha$ and its inverse respectively [13]. Although both the edges belong to different locally equivalence class, we observe from table 1 that they possess the same set of $s_l$. Further, it is also interesting to note that $s_l = 1/2$ for all the gates lie on the edge $A_2A_3$. Within the Weyl chamber the perfect entanglers form a polyhedron $LMNPQA_2$ with four edges $LQ$, $LM$, $A_2M$, and $A_2Q$ lie in the base (Fig. 1). We may note that the gates lie in the base are symmetric about the line $LA_2$. Therefore, the edges $LQ$ and $LM$ are locally equivalent to each other, possessing same set of $s_l$ (table 2). Similarly the edges $A_2M$ and $A_2Q$ are also locally equivalent to each other, hence share their coefficients $s_l$. We also note that the edges $QP$ and $MN$ possess same set of $s_l$, though the edges are not locally equivalent to each other.

Table. 2 Schmidt coefficients for the edges of polyhedron

| Edge | $[c_1, c_2, c_3]$ | Range of parameter | $s_1$ | $s_2$ | $s_3$ | $s_4$ |
|---|---|---|---|---|---|---|
| LQ | $\left[\frac{\pi}{2}-\theta, \theta, 0\right]$ | $0 \leq \theta \leq \frac{\pi}{4}$ | $\frac{1}{\sqrt{2}}[\cos^2(\theta/2) + (\sin\theta)/2]$ | $\frac{1}{\sqrt{2}}[\cos^2(\theta/2) - (\sin\theta)/2]$ | $\frac{1}{\sqrt{2}}[\sin^2(\theta/2) + (\sin\theta)/2]$ | $\frac{1}{\sqrt{2}}[(\sin\theta)/2 - \sin^2(\theta/2)]$ |
| LM | $\left[\frac{\pi}{2}+\theta, \theta, 0\right]$ | $0 \leq \theta \leq \frac{\pi}{4}$ | $\frac{1}{\sqrt{2}}[\cos^2(\theta/2) - (\sin\theta)/2]$ | $\frac{1}{\sqrt{2}}[\cos^2(\theta/2) + (\sin\theta)/2]$ | $\frac{1}{\sqrt{2}}[(\sin\theta)/2 - \sin^2(\theta/2)]$ | $\frac{1}{\sqrt{2}}[\sin^2(\theta/2) + (\sin\theta)/2]$ |
| $A_2M$ | $\left[\frac{\pi}{2}+\phi, \frac{\pi}{2}-\phi, 0\right]$ | $0 \leq \phi \leq \frac{\pi}{4}$ | $\frac{1}{2}\cos\phi$ | $\frac{1}{2}[1+\sin\phi]$ | $\frac{1}{2}[1-\sin\phi]$ | $\frac{1}{2}\cos\phi$ |
| $A_2Q$ | $\left[\frac{\pi}{2}-\phi, \frac{\pi}{2}-\phi, 0\right]$ | $0 \leq \phi \leq \frac{\pi}{4}$ | $\frac{1}{2}[1+\sin\phi]$ | $\frac{1}{2}\cos\phi$ | $\frac{1}{2}\cos\phi$ | $\frac{1}{2}[1-\sin\phi]$ |
| QP | $\left[\frac{\pi}{4}, \frac{\pi}{4}, \eta\right]$ | $0 \leq \eta \leq \frac{\pi}{4}$ | $\sqrt{\cos^4\left(\frac{\pi}{8}\right)\cos^2\left(\frac{\eta}{2}\right) + \sin^4\left(\frac{\pi}{8}\right)\sin^2\left(\frac{\eta}{2}\right)}$ | $\frac{1}{2\sqrt{2}}$ | $\frac{1}{2\sqrt{2}}$ | $\sqrt{\sin^4\left(\frac{\pi}{8}\right)\cos^2\left(\frac{\eta}{2}\right) + \cos^4\left(\frac{\pi}{8}\right)\sin^2\left(\frac{\eta}{2}\right)}$ |
| MN | $\left[\frac{3\pi}{4}, \frac{\pi}{4}, \eta\right]$ | $0 \leq \eta \leq \frac{\pi}{4}$ | $\frac{1}{2\sqrt{2}}$ | $\sqrt{\cos^4\left(\frac{\pi}{8}\right)\cos^2\left(\frac{\eta}{2}\right) + \sin^4\left(\frac{\pi}{8}\right)\sin^2\left(\frac{\eta}{2}\right)}$ | $\sqrt{\sin^4\left(\frac{\pi}{8}\right)\cos^2\left(\frac{\eta}{2}\right) + \cos^4\left(\frac{\pi}{8}\right)\sin^2\left(\frac{\eta}{2}\right)}$ | $\frac{1}{2\sqrt{2}}$ |
| PN | $\left[\frac{\pi}{4}+\eta, \frac{\pi}{4}, \frac{\pi}{4}\right]$ | $0 \leq \eta \leq \frac{\pi}{2}$ | $\sqrt{\cos^4\left(\frac{\pi}{8}\right)\cos^2\left(\frac{\pi}{8}+\frac{\eta}{2}\right) + \sin^4\left(\frac{\pi}{8}\right)\sin^2\left(\frac{\pi}{8}+\frac{\eta}{2}\right)}$ | $\sqrt{\sin^4\left(\frac{\pi}{8}\right)\cos^2\left(\frac{\pi}{8}+\frac{\eta}{2}\right) + \cos^4\left(\frac{\pi}{8}\right)\sin^2\left(\frac{\pi}{8}+\frac{\eta}{2}\right)}$ | $\frac{1}{2\sqrt{2}}$ | $\frac{1}{2\sqrt{2}}$ |
| LN | $\left[\frac{\pi}{2}+\theta, \theta, \theta\right]$ | $0 \leq \theta \leq \frac{\pi}{4}$ | $\frac{1}{2}\sqrt{1+\cos^2\theta - \sin 2\theta}$ | $\frac{1}{2}\sqrt{1+\cos^2\theta + \sin 2\theta}$ | $\frac{1}{2}\sin\theta$ | $\frac{1}{2}\sin\theta$ |
| $A_2P$ | $\left[\frac{\pi}{2}-\theta, \frac{\pi}{2}-\theta, \theta\right]$ | $0 \leq \theta \leq \frac{\pi}{4}$ | $\frac{1}{2}\sqrt{1+\sin^2\theta + \sin 2\theta}$ | $\frac{1}{2}\cos\theta$ | $\frac{1}{2}\cos\theta$ | $\frac{1}{2}\sqrt{1+\sin^2\theta - \sin 2\theta}$ |

It is already shown that the edge $OA_1$ represents controlled unitary gates possessing Schmidt number 2. In this case, the Schmidt strength $K_{Sch}$ reduces to binary entropy such that $0 \leq K_{Sch} \leq 1$ with CNOT possessing the maximal value. Figure 2 shows the Schmidt strength of $OA_1$ and $OA_2$ with respect to their parameter. Since the edges $OA_3$ and $A_1A_3$ possess same set of $s_l$, the measure $K_{Sch}$ assumes same form as shown in Fig. 3(a). Another edge lying in the base is $A_2A_1$ for which $K_{Sch}$ is shown in Fig. 3(b). From Fig. 2 and Fig. 3, we observe that for the five edges of Weyl chamber the measure $K_{Sch}$ is monotonic with the parameter. In other words, no two gates lying in a given edge possess the same Schmidt strength. It is known that two-qubit operators having four non-vanishing Schmidt coefficients with the same amplitude must be maximally entangled [8]. Here we observe that all the gates in edge $A_2A_3$ possess equal Schmidt coefficients of 1/2 and hence $K_{Sch} = 2$. In other words, all the gates in the edge $A_2A_3$ (which include Double-CNOT and SWAP) are maximally entangled. It is interesting to note from the notion of entanglement of an operator that SWAP is maximally entangled, though it does not produce any entanglement when acting on a state.

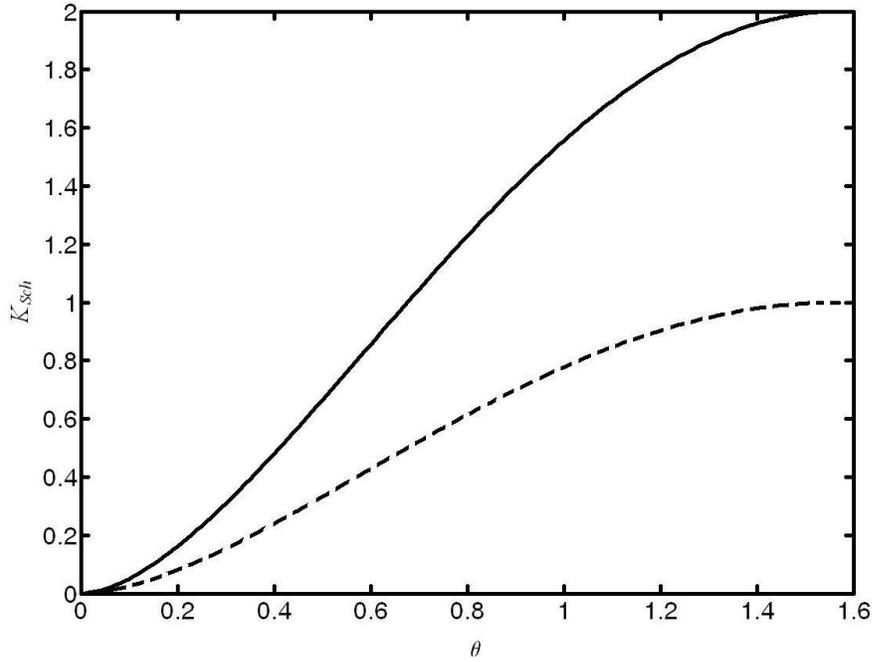

**Fig. 2** Schmidt strength of $OA_1$ (dashed line) and $OA_2$ (solid line). The other half of $OA_1$ is not shown on symmetry ground

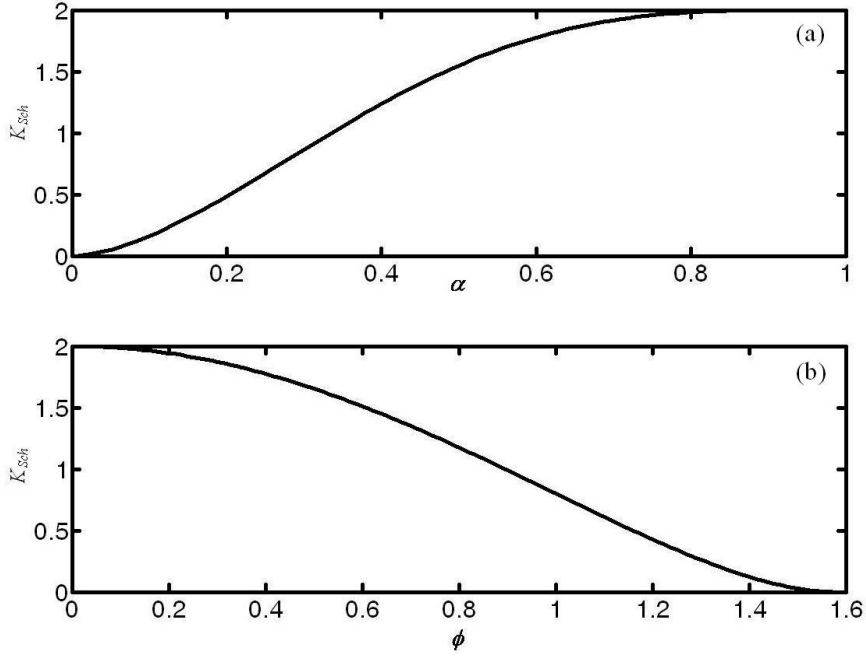

**Fig. 3 (a) & (b)** Schmidt strengths of $OA_3$ ($A_1A_3$) and $A_2A_1$ respectively

Since the polyhedron edges *LQ* and *LM* are locally equivalent to each other, they possess same form of Schmidt strength. Similarly, the edges $A_2M$ and $A_2Q$ also assume the same form of $K_{Sch}$ (Fig. 4). The same figure shows $K_{Sch}$ for *LN* and $A_2P$ as well. Figure 5(a) shows $K_{Sch}$ for *QP* which is same as that of *MN*. We note that the Schmidt strength for all the edges of polyhedron discussed so far are monotonic functions of their parameter. Therefore, no two gates lying in a given edge possess the same Schmidt strength. However, the edge *PN* is an exception for which $K_{Sch}$ is not a monotonic function of parameter. For this edge, the measure is symmetric about its midpoint (Fig. 5(b)). Further, we also observe that $1 \leq K_{Sch} \leq 2$ for all the edges of polyhedron.

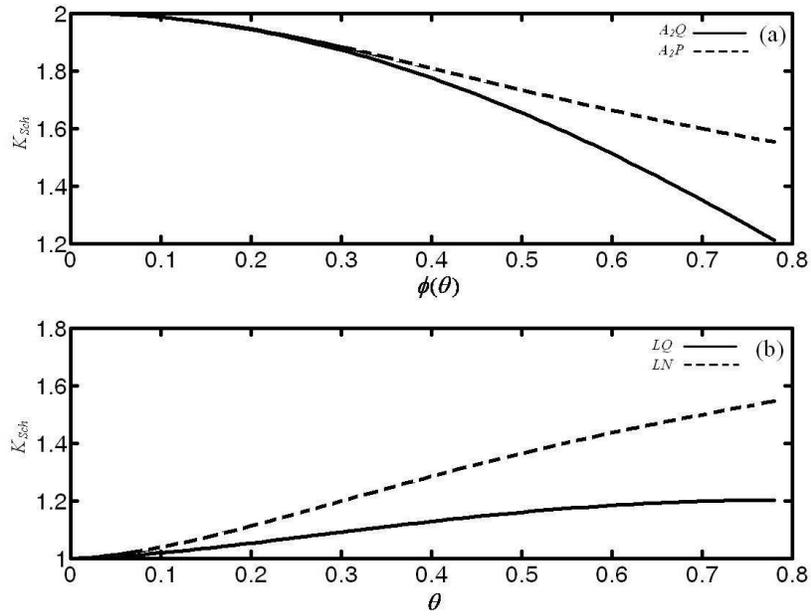

**Fig. 4 (a) & (b)** Schmidt strengths of $A_2Q$ ($A_2M$), $A_2P$ and $LQ$ ($LM$), $LN$ respectively

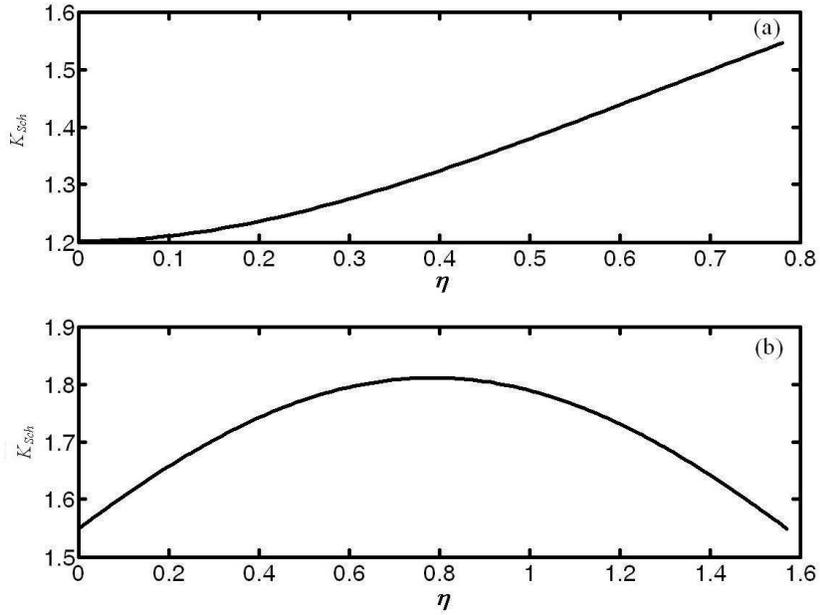

**Fig. 5 (a) & (b)** Schmidt strengths of $QP$ ($MN$) and $PN$ respectively

## 5 Conclusion

In this paper we have studied the operator-Schmidt decomposition of nonlocal two-qubit gates. It is shown that locally equivalent gates must necessarily possess *same* set of Schmidt coefficients $s_l$. However, the vice-versa is not true. Employing geometrical approach, it is shown that controlled unitary gates (one edge of the Weyl chamber) *only* have Schmidt number 2 and all other gates possess Schmidt number 4. Hence CNOT is the only perfect entangler possessing Schmidt number 2. Thus, our approach compliments the Schmidt number classification of two-qubit gates [6].

Further, all the six edges of Weyl chamber and nine edges of polyhedron are characterized using Schmidt coefficients $s_l$ and Schmidt strength $K_{Sch}$ – a measure of entanglement of two-qubit gate such that $0 \leq K_{Sch} \leq 2$. It is found that except one edge of the Weyl chamber and polyhedron, for every edge the gates possess unique Schmidt strength. In addition, it is found for all the edges of polyhedron that $1 \leq K_{Sch} \leq 2$. It remains to be checked if this range of $K_{Sch}$ is valid for all the perfect entanglers as well.

It is found that $K_{Sch} = 2$ for one edge ($A_2A_3$) of the Weyl chamber in which SWAP and Double-CNOT are the two members. In other words, all the gates lie in the edge $A_2A_3$ are maximally entangled. It is worth mentioning that the notion of entanglement of a gate is radically different from entanglement production of a state upon action of the gate. For instance, SWAP is known to be a maximally entangled gate though it does not produce any entanglement on a state. D. Collins *et al*. introduced a framework to measure the nonlocal content of a gate, in terms of resources (entangled and classical bits) required to implement the gate through double teleportation. Within this framework, SWAP and Double-CNOT are known to be maximally nonlocal gates. It is then interesting to check if all the gates in the edge $A_2A_3$ are also maximally nonlocal.

**Acknowledgements**   S. B acknowledges Council of Scientific and Industrial Research, India for the financial support in the form of senior research fellowship.